\documentstyle[twoside,fleqn,espcrc2]{article}

\input epsf
\epsfverbosetrue

 \title{\vspace*{-15mm}\begin{flushright} {\normalsize NORDITA-1999/50 HE} \end{flushright}
\vspace*{-5mm}  
Interactions of heavy-light mesons\thanks{Presented by P. Pennanen, 
 {\tt petrus@hip.fi}}}

 \author{{\bf UKQCD} Collaboration: P.~Pennanen\address{Nordita, Blegdamsvej 17, 2100 Copenhagen \O, Denmark}, 
 C.~Michael\address{Theoretical Physics Division, Dept. of Math.
 Sciences,
 University of Liverpool, Liverpool, UK}$^\dagger$ and A.M.~Green\address{Dept. of Phys. and Helsinki Inst. of Phys.,
 P.O. Box 9, FIN-00014 University of Helsinki,
 Finland}\thanks{e-mails: {\tt cmi@liv.ac.uk, anthony.green@helsinki.fi}}}

\begin{document}

 \begin{abstract}
 The potential between static-light mesons forming a meson-meson or a 
 meson-antimeson system is calculated in quenched and unquenched SU(3) gauge 
 theory. We use the Sheikholeslami-Wohlert 
 action and statistical estimators of light quark propagators with maximal 
 variance reduction. The dependence of the potentials on the light
 quark spin and isospin and the effect of meson exchange is investigated. Our 
 main motivation is exploration of bound states of two mesons and string 
 breaking. The latter also involves the
 two-quark potential and the correlation between two-quark and two-meson states.
 \end{abstract}

 \noindent
 \maketitle

We have been working on lattice calculations of multiquark systems in order
to understand their properties from first principles. Here ``multi'' means that
the system can be decomposed into more than one colour singlet, the simplest
case being four quarks. 
Previously, we have studied  four quarks in the static approximation 
in SU(2), getting energies for a general set of geometries which have been 
reproduced by a model based on two-body potentials and multiquark interaction 
terms~\cite{gre:98}. We have also looked at the flux distribution 
corresponding to the 
binding energy~\cite{pen:98}. These results for the static case are discussed 
in last year's proceedings. 

In the current work only two of the quarks are static and SU(3) is used,
also unquenched.
The binding of this system of two heavy-light mesons is studied as a function 
of the heavy quark separation. Different cases of spin and isospin of the
light quarks are measured and the effect from meson exchange is extracted
from the potentials. The model for energies of static quarks is extended for
this more dynamic system as discussed in Ref.~\cite{gre:99} and by A.M. Green 
in these
proceedings. The meson-antimeson case with light quark isospin $I_q=0$ 
is relevant to string breaking.

In the future we plan to explore breaking of an excited string,
which is interesting from the point of view of hybrid meson phenomenlogy.

\vspace{-0.2cm}

\section{TWO HEAVY-LIGHT MESONS}

Experimental candidates for bound states of four quarks (of which two are
antiquarks) lie close to meson-antimeson thresholds. In our case e.g. a heavy 
$\Upsilon$ particle 
 could be a $B^*\bar{B}^*$ system. Systems with heavy quarks should be more
 easily bound as the repulsive kinetic energy is smaller with the attractive
 potential being flavour independent. The binding of four-quark systems has
 been studied e.g. in bag, string-flip and deuson models, states with 
 two $b$ quarks being stable in most models.

 In the lattice calculation we measure two diagrams for both meson-meson
 and meson-antimeson cases -- see Fig. 1. One of them is the unconnected one without
 light quark interchange and the other connected, where the light quarks hop
 from one meson to another. The light quark mass we use is approximately
 that of $s$. Because of the static approximation for the heavy quarks
 their spin and isospin decouples making the pseudoscalar $B$
 and the vector $B^*$ degenerate, while physically they have a 46 MeV (1\%) 
 separation. We call this degenerate set ${\cal B}$. For the two-meson system 
 combinations of $B$ and $B^*$ do have
 different energies in our case. We measure wavefunctions symmetric under
 interchange of the mesons with the light quark spin and isospin being 
 singlet or triplet; these then couple to $B, B^*$ combinations~\cite{mic:99}. 

 \begin{figure}[h]
 \vspace{-2cm}
 \begin{center}
 \epsfxsize=250pt\epsfbox{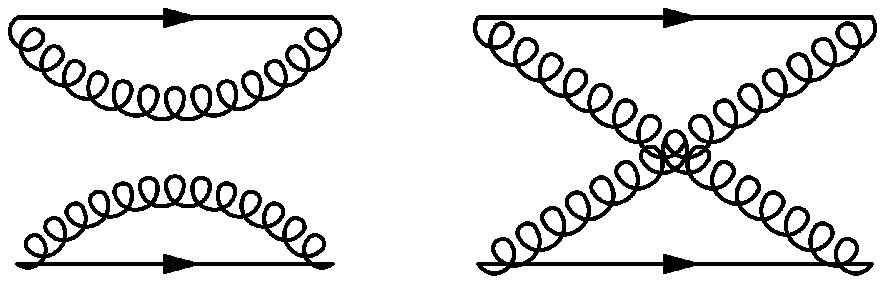}
 \vspace{-10.9cm}
 \epsfxsize=250pt\epsfbox{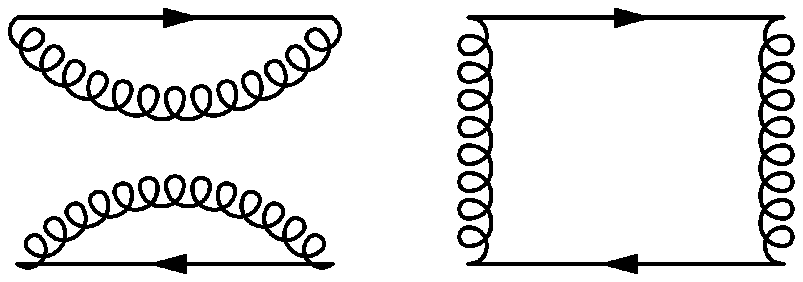}
 \end{center}
 \vspace{-10.9cm}
 \caption{Diagrams for ${\cal B}{\cal B}$ and ${\cal B}\bar{\cal B}$ systems.}
 \vspace{-0.7cm}
 \end{figure}

 We obtain estimates of light quark propagators using maximally variance reduced
 pseudo-fermionic ensembles~\cite{mic:98}, 24 of which are generated for each 
 gauge configuration. A second nested Monte Carlo calculation is then performed
 for the pseudofermions of each gauge configuration. The variance reduction
 requires ``halving'' of the lattice into two separate regions with the 
 propagators going from one region to the other. Thus the connected diagram
 for ${\cal B}\bar{\cal B}$ and the cross correlator between $Q\bar{Q}$ and
 ${\cal B}\bar{\cal B}$ need to have one of the normally 
spatial
 axes as the temporal axis, introducing considerable technical complications. The parameters of
 our calculation are $\beta=5.2, \ C_{SW}=1.76, \ a\approx 0.14 \ {\rm fm}, \ 
 M_{PS}/M_V=0.72$ for unquenched~\cite{all:98} and $\beta=5.7, \ C_{SW}=1.57, \ 
 a\approx 0.17 \ {\rm fm}, \ M_{PS}/M_V=0.65$ for the quenched
 case with a $16^3\times24$ lattice being used for both. We have 20 and 54
 gauge configurations for quenched and unquenched respectively; with 
 pseudofermionic fields these take some 60 GB of diskspace. A variational 
 basis of local and fuzzed mesonic operators is used, diagonalization of which
 maximizes overlap with the ground state of the system. 

 We measure the different spin and isospin components as discussed in 
 Ref.~\cite{mic:99}. For the ${\cal B}\bar{\cal B}$ case $I_q=1$ has only
 the unconnected diagram, whereas $I_q=0$ has the connected one 
 subtracted. 

 \vspace{-0.2cm}

 \section{RESULTS}

 The raw correlators show that for the ${\cal B}{\cal B}$ system the unconnected
 diagram is much noisier and does not contribute to the binding at $R>1$.
 The connected diagram, on the other hand, gives a small binding for larger
 $R$ and also contributes to the observables where the spin of the light quarks
 changes. 

At $R=0$ the heavy quarks are at the same point and the ${\cal B}{\cal B}$ 
case looks like a baryon with an antitriplet string. We can compare to previously 
measured~\cite{mic:98} energies of the $\Lambda_b, \ \Sigma_b$ baryons for 
$I_q,S_q=(0,0)$ and $I_q,S_q=(1,1)$ respectively, 
finding excellent agreement~\cite{mic:99}. States with a sextet string
$(0,1;\ 1,0)$ lie higher. The ${\cal B}\bar{\cal B}$ singlet
at $R=0$ looks like a pion, and we find agreement with the energy of a
pion with non-zero momentum. 

The meson-meson potentials are shown in 
Fig. 2. The $I_q, S_q = (1,1)$ case is similar to $(0,0)$ but less
bound; the level ordering observed at $R=0$ is retained and the attraction
disappears earlier. For the $(1,0)$ a remnant
of the sextet string makes the small-$R$ potential repulsive. The $(0,1)$ at 
$R=0$ is attractive for unquenched and repulsive for quenched; this is the only
qualitative ($\approx 2.5\sigma$) difference between the quenched and unquenched 
results visible in our data. Both $(0,1)$ and
$(1,0)$ seem to have attraction at $r\approx 0.3$ fm, which is a meson 
exchange effect as opposed to the small distance behaviour governed by 
gluonic effects. From a crude two-body Schr\"odinger approach using these 
potentials we expect binding for all of these cases except perhaps $(1,0)$. 

 \begin{figure}[h]
 \vspace{-0.3cm}
 \begin{center}
 \epsfxsize=210pt\epsfbox{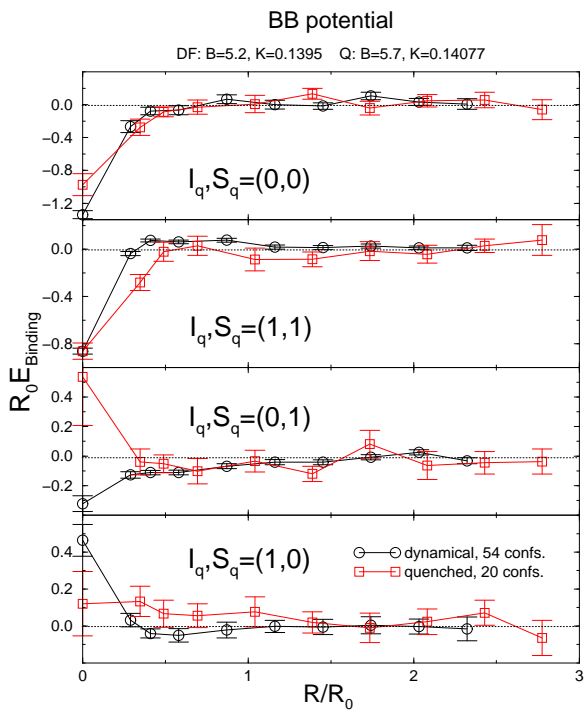}
 \end{center}
 \vspace{-1.8cm}
 \caption{${\cal B}{\cal B}$ potentials. Here $2M_BR_0=4.98(1), \ 5.83(3)$ for quenched, unquenched respectively and $aR_0=0.49$ fm.}
 \vspace{-0.8cm}
 \end{figure}

 The meson-antimeson potential for $I_q, S_q = (1,0)$ is shown in Fig. 3, the
 $(1,1)$ case being similar. For both of these a $Q\bar{Q}+q\bar{q}$ state 
 with the same quantum number is lighter for small $R$. For $(1,0); (1,1)$ the 
 relevant energies are $V(R)+\pi$ and $V(R)+\rho$ respectively, where 
 $V(R)$ is the static $Q\bar{Q}$ potential. An estimate of
 $V(R)+\pi$ is included in Fig. 3. 

 \begin{figure}[h]
 \vspace{-1.15cm}
 \begin{center}
 \epsfxsize=210pt\epsfbox{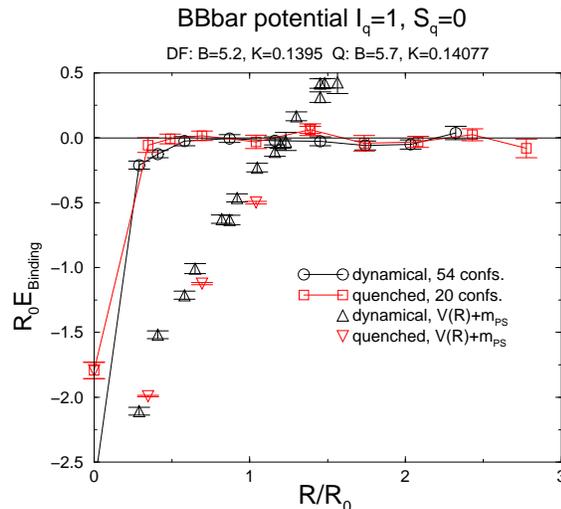}
 \end{center}
 \vspace{-2cm}
 \caption{${\cal B}\bar{\cal B}$ potentials. For small $R$ a state with a 
$Q\bar{Q}$ and a pion should be lighter.}
 \vspace{-0.6cm}
 \end{figure}

 The contribution from meson exchange can be examined e.g. by looking at
 the crossed diagram measured in the quenched approximation at fixed $T$ as
 a function of $R$. The $BB \rightarrow BB$ case should have a contribution from
 $\rho$ exchange, and we indeed find agreement by using the previously
 measured $m_\rho$ and normalizing by hand. For $BB^* \rightarrow B^*B$ 
 we should have a contribution from $\pi$ exchange. In this case we can use 
 a recent determination of the $BB^*\pi$ coupling~\cite{div:98}, the 
 experimental decay constant and our $m_\pi$. In the one-$\pi$ exchange formula
 everything is thus known, and we find excellent quantitative agreement for
 $R\ge 0.5$ fm. This is strong support for deuson models~\cite{tor:94} in this 
 distance range.

 \section{STRING BREAKING}

 In our quenched and unquenched calculations the ground state 
 ${\cal B}\bar{\cal B}$ and $Q\bar{Q}$ potentials cross at $r\approx 1.2$ fm. 
 We are investigating the breaking of the $Q\bar{Q}$ string by using a variational approach
 similar to that used in Higgs models by several groups. The cross 
 correlator between two-meson and two-quark states allows us to study their
 mixing also in the quenched theory -- in the unquenched case additional fermion
 bubbles induce corrections. The quenched mixing matrix element can then be 
 used to estimate the splitting of energy levels at the string breaking point,
 even though no actual splitting occurs with quenching. 
 With an unquenched calculation the energy splitting can be studied directly 
 using the full variational approach.

 One might think that an excited string would break at a smaller distance than 
 the ground state. This is not necessarily the case, as e.g. the first
 excited state has $J_z=1$ with quark separation along $z$ and only breaks
 into mesons ${\cal B}_L\bar{\cal B}_{L'}$ with $L+L'>0$. In general it is
 an open question if a state with particular quantum numbers has the lowest
 energy at a particular heavy quark separation as a) a hybrid $Q\bar{Q}$ meson 
 with excited glue, b) a ground state 
 $Q\bar{Q}$ meson and a $q\bar{q}$ meson or c) two heavy-light mesons. These 
 energy levels and their mixing can 
 be studied on the lattice with our techniques. 

 \vspace{-0.2cm}

\newcommand{\href}[2]{#2}\begingroup\raggedright\endgroup

 
\end{document}